\documentclass[sigconf]{acmart}
\usepackage{xcolor}
\usepackage[ruled,linesnumbered]{algorithm2e}
\usepackage{graphicx}
\usepackage{subfig}
\usepackage{threeparttable}
\usepackage{booktabs}
\usepackage{multirow}
\usepackage{caption}
\usepackage{url}
\usepackage{diagbox}
\usepackage{svg}
\usepackage{enumitem}
\usepackage{algorithmic}
\usepackage{float}
\theoremstyle{definition}
\newtheorem{definition}{Definition}[section]

\newcommand{\cheng}{\color{black}}

\begin{document}

\title{Data Watermarking for Sequential Recommender Systems}

\author{Sixiao Zhang}
\affiliation{%
  \institution{Nanyang Technological University}
  \city{Singapore}
  \country{Singapore}
}
\email{sixiao001@e.ntu.edu.sg}

\author{Cheng Long}
\authornote{Co-corresponding authors.}
\affiliation{%
  \institution{Nanyang Technological University}
  \city{Singapore}
  \country{Singapore}
}
\email{c.long@ntu.edu.sg}

\author{Wei Yuan}
\affiliation{%
  \institution{The University of Queensland}
  \city{Brisbane}
  \country{Australia}
  }
\email{w.yuan@uq.edu.au}

\author{Hongxu Chen}
\affiliation{%
  \institution{The University of Queensland}
  \city{Brisbane}
  \country{Australia}
  }
\email{hongxu.chen@uq.edu.au}

\author{Hongzhi Yin}
\authornotemark[1]
\affiliation{%
  \institution{The University of Queensland}
  \city{Brisbane}
  \country{Australia}
}
\email{h.yin1@uq.edu.au}

\begin{abstract}
In the era of large foundation models, data has become a crucial component in building high-performance AI systems. As the demand for high-quality and large-scale data continues to rise, data copyright protection is attracting increasing attention. In this work, we explore the problem of data watermarking for sequential recommender systems, where a watermark is embedded into the target dataset and can be detected in models trained on that dataset. We focus on two settings: dataset watermarking, which protects the ownership of the entire dataset, and user watermarking, which safeguards the data of individual users. We present a method named Dataset Watermarking for Recommender Systems (DWRS) to address them. We define the watermark as a sequence of consecutive items inserted into normal users' interaction sequences. We define a Receptive Field (RF) to guide the inserting process to facilitate the memorization of the watermark. Extensive experiments on five representative sequential recommendation models and three benchmark datasets demonstrate the effectiveness of DWRS in protecting data copyright while preserving model utility.
\end{abstract}

%
%



\maketitle
\sloppy
\section{Introduction}
Recommender systems \cite{rendle2012bpr,zhang2019deep,he2020lightgcn,wu2022graph,long2023decentralized} are a representative real-world application of machine learning systems. They are deployed on various platforms that are closely related to people's daily lives, such as e-commerce apps \cite{schafer2001commerce,yuan2024hide}, video sharing sites \cite{davidson2010youtube}, social platforms \cite{fan2019graph}, etc. Massive high-quality data is necessary to train a reliable and powerful recommender system. Such data may be partially obtained from open-source datasets and third-party providers. Since such data contains huge commercial values and potential user private information, malicious clients may conduct illegal usage and unauthorized redistribution as well as other prohibited behaviors on the obtained data, thus causing data safety and privacy issues for the data owners \cite{himeur2022latest,zhang2023comprehensive,yuan2023interaction,ge2024survey,zhang2024defense,yuan2025ptf}. Therefore, data owners need a promising method to trace down the usage of their data and claim ownership.

\begin{figure}[t]
\centering
    \includegraphics[width=0.4\textwidth]{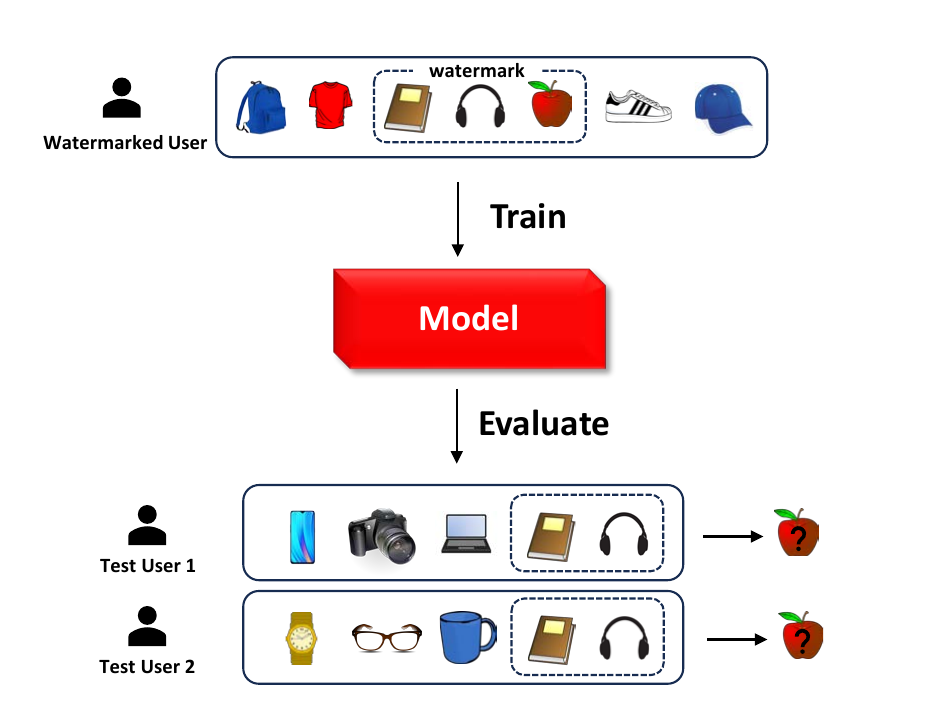}
\caption{An illustration of data watermarking for recommender systems. A watermark sequence (books, headphones, apples) is inserted into a training sequence. The model would memorize the watermark by predicting apples for test queries ended with books and headphones.}
\label{fig:illustration}
\vspace{-2em}
\end{figure}

Data watermarking has been studied as an effective solution to the data ownership protection problem in computer vision \cite{sablayrolles2020radioactive,hu2022membership,li2022untargeted}, natural language processing \cite{liu2023watermarking}, and graphs \cite{zhang2021backdoor}. A specific watermark pattern is defined and added to the benign data by the data owner. A model trained with the watermarked dataset can be queried with watermarked data samples to acquire a dedicated output that normally would not be produced by a non-watermarked model. In this way, the data owner can claim that this model has been trained on their data without authorization. For example, images are often watermarked by pixel-level noise such as a patch of white or black pixels \cite{hu2022membership,li2023black}; texts are often watermarked by synonyms or additional phrases \cite{liu2023watermarking,wei2024proving,li2023functionmarker}; graphs are often watermarked by inserting subgraphs or nodes \cite{zhang2021backdoor,xu2022poster}. However, solutions in these domains cannot be transferred to recommender systems because of the distinct data structures and downstream tasks. In recommender systems, the training data involves a set of users and a set of items, where each user has interactions with a subset of items. The task is to recommend top-k items that a user is most likely to adopt, given the user's interaction history.

Unique challenges exist in watermarking user-item interaction data for recommender systems. First, as a new research problem, we lack a clear definition of data watermarking for recommender systems and the criteria that a good watermark should meet, which hinders the development of a systematic and universal watermarking framework. A straightforward solution is to insert a special interaction pattern into benign data, so that the model would memorize the pattern after training. An illustration is shown in \autoref{fig:illustration}, where we define the watermark as a sequence of three consecutive items including books, headphones, and apples, then insert it into benign user's interaction data. After training, the model is expected to memorize the watermark so that we can extract it by appending books and headphones to the end of test queries and observe a highly confident prediction of apples. However, some specific design aspects remain unclear, such as the choice of watermark items, the selection of users to watermark, and the positions for inserting the watermark. Second, apart from watermarking the entire dataset, one may be interested in watermarking the interaction data of a single user in order to protect the data privacy of an individual customer, which is a more challenging.

\begin{figure}[t]
\centering
    \includegraphics[width=0.4\textwidth]{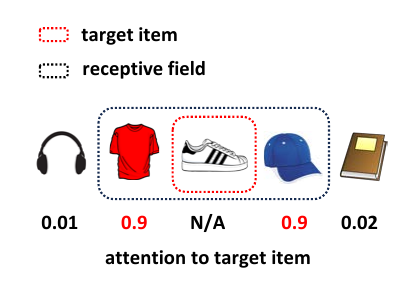}
\vspace{-2.5em}
\caption{An illustration of the receptive field. For the target item, the shirt and the hat have large attention values to it, while the headphone and the book contribute little. Therefore, the receptive field of the target item includes its previous item and subsequent item.}
\label{fig:introduction}
\vspace{-2em}
\end{figure}

To this end, we study the problems of dataset watermarking and user watermarking for recommender systems. We present a comprehensive analysis and definitions of the problems and introduce our proposed methods. We name our method as \textbf{D}ata \textbf{W}atermarking for \textbf{R}ecommender \textbf{S}ystems (DWRS) with its two variants, DWRS-D for dataset watermarking and DWRS-U for user watermarking. Since the evaluation of the watermark requires the model to be able to process unseen queries, we focus on the most popular inductive recommendation task: the sequential recommendation task. We define the watermark as a sequence of consecutive items. For DWRS-D, we define a \textbf{receptive field} for each dataset. It is defined to be the range of items that contribute significantly to the embedding of a target item. An illustration of the receptive field is shown in \autoref{fig:introduction}. Specifically, in each iteration, we insert the watermark into a benign user interaction sequence, and record the set of items (e.g. shirt and hat in \autoref{fig:introduction}) within the receptive field of the watermark. These items are named as \textbf{filler items}. In the next iteration, we insert the watermark into a new user interaction sequence, such that its receptive field does not contain any filler items from previous watermarked user interaction sequences. In this way, the model would be less likely to establish correlations between filler items and the watermark, but instead learn a strong correlation among items within the watermark, thus facilitating the memorization. For DWRS-U, we insert the watermark into the target user's interaction sequence. Since we can only insert the watermark into a single user interaction sequence in user watermarking, the watermark would be much longer compared with DWRS-D to achieve a high success rate. To reduce the watermark length and facilitate the memorization of the watermark, we compute the average item popularity of each subsequence in the target user's interaction sequence. We call the subsequence which has the smallest average item popularity as the most \textbf{unpopular} subsequence in the target sequence, and insert the watermark right before it, so that the unpopular subsequence would help provide additional overfitting capacity in memorizing the watermark to compensate for the reduced length, as the model would try to establish correlations between the watermark and the unpopular subsequence. We conducted extensive experiments on five representative sequential recommender systems and three popular datasets. DWRS achieves high watermark validity while preserving the model utility. It is stealthy due to the short watermark length and small insert ratio. It is also robust against three watermark removal attacks, including finetuning, distillation, and sequential rule mining.
Our code is available at \url{https://github.com/RinneSz/DWRS}.

We summarize our contributions as follows:
\begin{itemize}
    \item We define the problems and constraints of dataset watermarking and user watermarking for sequential recommender systems. To the best of our knowledge, we are the first to study data watermarking for recommender systems.
    \item We propose DWRS with its two variants DWRS-D and DWRS-U targeting at dataset watermarking and user watermarking respectively. DWRS inserts a watermark into the dataset so that the watermark can be detected on models trained on the dataset, thus making it possible for the data owner to claim the ownership and avoid unauthorized use of the data.
    \item Extensive experiments demonstrate the effectiveness of DWRS in injecting a valid watermark while preserving model utility. It is also robust against finetuning, distillation, and sequential rule mining.
\end{itemize}

\section{Related Works}
We review prior works on data watermarking in computer vision, natural language processing, and graphs.

In computer vision, backdoor attacks are frequently used as the data watermark. Sablayrolles et al. \cite{sablayrolles2020radioactive} proposed to watermark images by adding a vector shift on the extracted image features in the hidden layer as the watermark. Then they backpropagate the watermark to compute the specific pixel changes. Hu et al. \cite{hu2022membership} and Li et al. \cite{li2023black} proposed to backdoor images using BadNets attack, where a backdoor trigger overwrites some pixels of the benign images. Li et al. \cite{li2022untargeted} proposed to alter the training image itself but not the label, so that the posterior probability of the ground-truth class is significantly different between the clean image and the backdoored image. Tang et al. \cite{tang2023did} proposed a clean-label backdoor watermarking. They randomly select some examples which belong to a particular class and add a trigger to them, so that the model will learn a strong correlation between the trigger and that class. Guo et al. \cite{guo2024domain} identify hard samples that will be misclassified by the benign model, and then add watermarks to them so that the backdoored model will correctly classify them into the ground-truth class.

In natural language processing, the most popular data watermarking methods are conducted by word-level and sentence-level backdoor attacks. Liu et al. \cite{liu2023watermarking} proposed to watermark text data for text classification based on existing backdoor attacks on LLMs. Wei et al. \cite{wei2024proving} proposed to watermark text data used for pretraining LLMs. They proposed two watermarks. One is to generate random sequences and append them to the end of each document. Another is to replace some tokens with their Unicode lookalikes. Li et al. \cite{li2023functionmarker} proposed to watermark text data in LLMs by prompts. They design watermarks as prompt questions and answers.

In graphs, data watermarking is conducted by inserting nodes or subgraphs, as well as changing the features. Zhang et al. \cite{zhang2021backdoor} proposed to watermark graphs for graph classification by using a random subgraph trigger to replace some of the nodes in a target graph and then change the label of the graph into a predefined class. Xu et al. \cite{xu2022poster} proposed a clean-label backdoor attack on graph classification. The idea is to insert an ER random graph to a specific class so that the model will establish a strong correlation between the trigger and the class. Xing et al. \cite{xing2023clean} proposed a clean-label backdoor attack on node classification. They use node features as the trigger to establish correlations with a class.

However, due to fundamental differences in data structures and task-specific requirements, existing watermarking methods cannot be directly applied to protecting data in recommender systems. This highlights the urgent need for specialized watermarking techniques tailored to this domain. Recently, Zhang et al. \cite{zhang2024watermarking} proposed a model watermarking approach for recommender systems. However, their method is specifically designed to protect model copyright, operating under the assumption that the attacker lacks access to the dataset. Their approach defines the watermark as an independent sequence that appears multiple times within the dataset. Consequently, if applied to data watermarking, the watermark would become easily identifiable and susceptible to removal.
\section{Dataset Watermarking}

\subsection{Problem Definition}
\label{sec:def 1}
A dataset watermark is a unique identifier that protects the copyright of the dataset. It is composed of a watermark body $x_{wm}$ and a target response $y_{wm}$. Given any data sample $x_{i}$, if $x_{wm}$ is embedded into $x_{i}$, then the model should output $y_{wm}$. For example, in image classification, $x_{wm}$ can be a patch of white or black pixels, and $y_{wm}$ can be the class of dog. Specifically, a good dataset watermark should satisfy the following constraints at the inference stage:
\begin{equation}
\label{eq:def 1}
    y_{wm} = f_{wm}(x\oplus x_{wm})
\end{equation}
\begin{equation}
\label{eq:def 2}
    y_{wm} \neq f(x\oplus x_{wm}), \text{ if } y_{wm}\neq f(x).
\end{equation}
where $x$ is the input, $f_{wm}$ is the watermarked model which is trained on the watermarked dataset, and $f$ is the oracle model that is trained on the clean dataset.  The operation $\oplus$ denotes the specific fusion operation for different tasks, such as the pixel overwriting in images, as well as token concatenation in texts. The first constraint ensures that the watermark can produce the desired response. The second constraint ensures that the watermarked model can be well distinguished from the oracle model.

While it is straightforward to define $x_{wm}$ and $y_{wm}$ in traditional classification tasks, it is however non-trivial in the recommendation task. In this work, we consider the sequential recommendation scenario, which is one of the most representative inductive recommendation tasks, because the evaluation of the watermark requires the model to be inductive, i.e., capable of dealing with unseen queries. Given a set of observed users $\mathcal{U}=\{u_{1},u_{2},...,u_{|\mathcal{U}|}\}$ and a set of items $\mathcal{I}=\{i_{1},i_{2},...,i_{|\mathcal{I}|}\}$, where each user $u_{j}$ has chronologically interacted with a set of items $S_{j}=\{i^{j}_{1},i^{j}_{2},...,i^{j}_{|S_{j}|}\}$, the task of sequential recommendation is to predict the next item $i^{j}_{|S_{j}|+1}$ that a user $u_{j}$ may interact with, given his or her interaction history $S_{j}$.

We outline several key properties essential for designing an effective dataset watermark for recommender systems:

\begin{itemize}
\item \textbf{Universality.} As defined in \autoref{eq:def 1}, the watermark should be applicable to every data sample, meaning it can be embedded into any sample while still eliciting the desired response from the model.
\item \textbf{Unnoticeability.} The watermark should be difficult to notice. This requires it to be short, infrequent, and minimally disruptive to model utility.
\item \textbf{Discriminability.} There should be a clear distinction between the watermark success rate of the watermarked model and that of the oracle model. Ideally, the watermarked model should achieve a 100\% watermark success rate, while the oracle model should have a 0\% success rate.
\item \textbf{Robustness.} The watermark should be resilient to removal attacks. A robust watermark remains detectable after an attack or significantly degrades the attacker's model utility.
\end{itemize}

Next, we present our proposed method, DWRS-D, and demonstrate how it satisfies these properties.

\begin{algorithm}[ht]
\caption{DWRS-D}
\label{alg:DWRS-D}
\begin{algorithmic}[1]
\REQUIRE
User set $\mathcal{U}$, user interaction sequence set $\mathcal{S}$, insert \\ratio $p$, watermark sequence $S_{wm}$, receptive field $RF$;
\ENSURE
Watermarked interaction sequence set $\mathcal{S}'$;
\STATE $\mathcal{S}'=\mathcal{S}$;
\STATE Initialize set of filler items $\mathcal{I}_{filler} \gets \emptyset$;
\STATE Initialize set of watermarked user sequences $\mathcal{U}_{wm} \gets \emptyset$;
\WHILE{$|\mathcal{U}_{wm}|<p\cdot|\mathcal{U}|$} \label{aaa}
    \STATE Randomly sample a user $u$ from $\mathcal{U}$ without replacement;
    \STATE $S_{u}\gets$ user $u$'s interaction sequence;
    \STATE Randomly sample a position $pos$ from $S_{u}$;
    \WHILE{$RF(pos)\cup \mathcal{I}_{filler} \neq \emptyset$}
        \STATE Resample $pos$ without replacement. If all positions have already been sampled, jump to line \autoref{aaa};
    \ENDWHILE
    \STATE $\mathcal{U}_{wm}=\mathcal{U}_{wm}\cup \{u\}$;
    \STATE $\mathcal{I}_{filler}=\mathcal{I}_{filler}\cup RF(pos)$;
    \STATE $S_{u}'\gets$Insert $S_{wm}$ into position $pos$ of $S_{u}$;
    \STATE Replace $S_{u}$ with $S_{u}'$ in $\mathcal{S}'$;
\ENDWHILE
\end{algorithmic}
\end{algorithm}

\begin{figure*}[t]
\centering
     \subfloat[ML-1M on SASRec]{\includegraphics[width=0.25\linewidth]{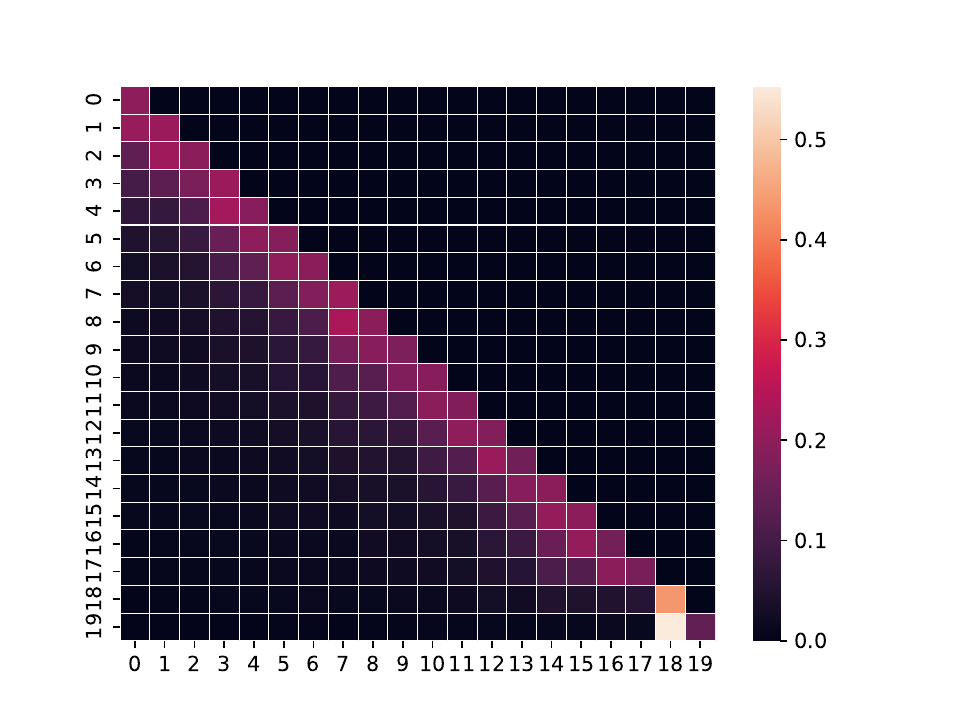}\label{fig:ml1m sas}}
	\subfloat[ML-1M on Bert4Rec]{\includegraphics[width=0.25\linewidth]{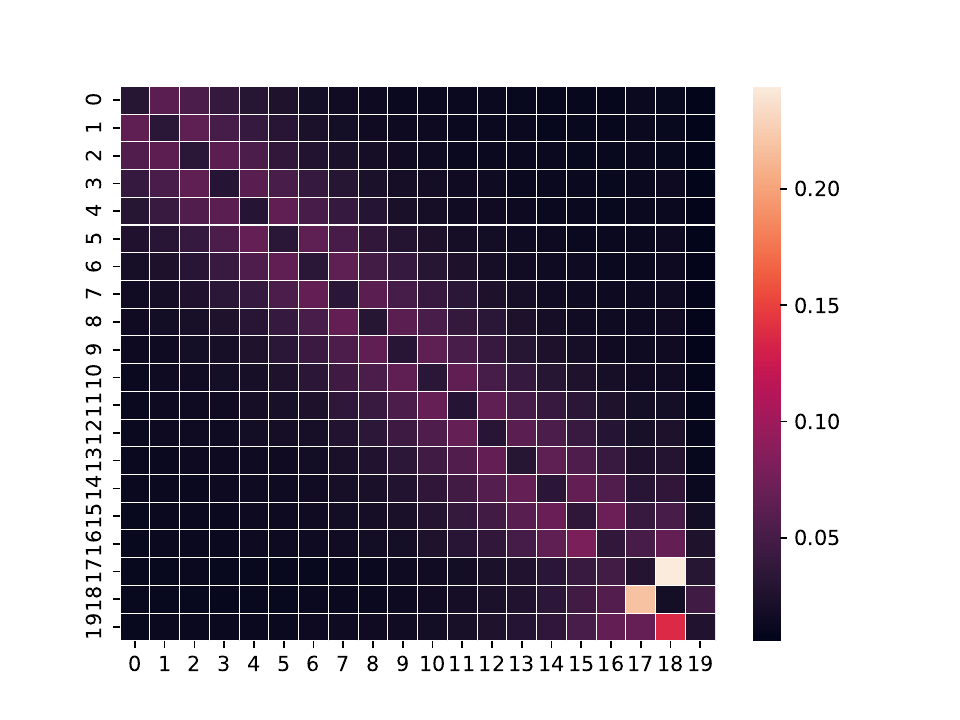}\label{fig:ml1m bert}}
    \subfloat[Beauty on SASRec]{\includegraphics[width=0.25\linewidth]{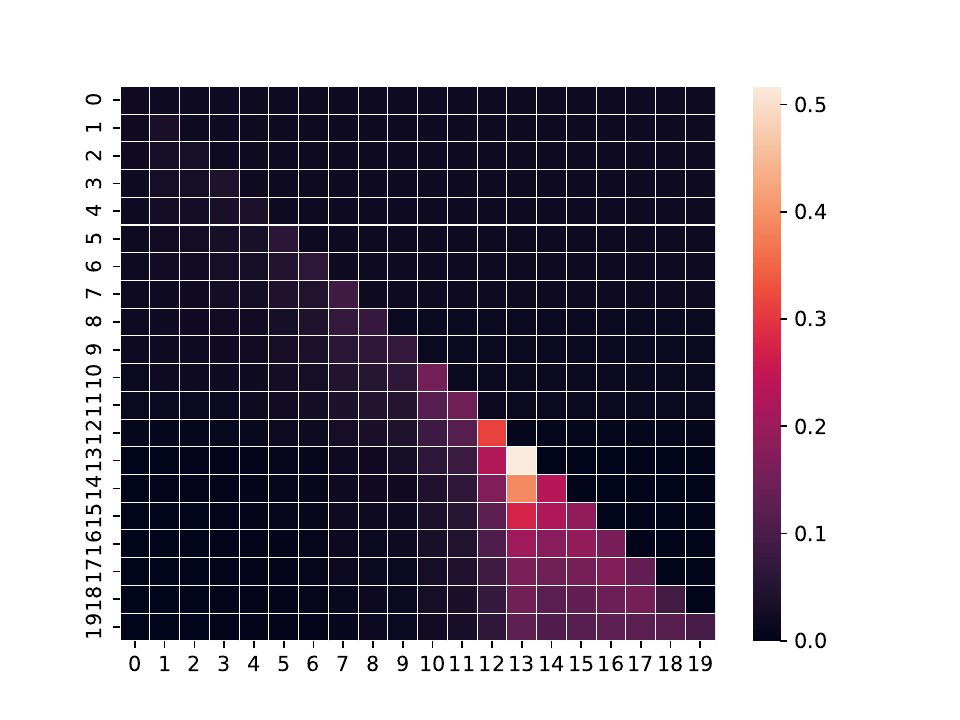}\label{fig:beauty sas}}
	\subfloat[Beauty on Bert4Rec]{\includegraphics[width=0.25\linewidth]{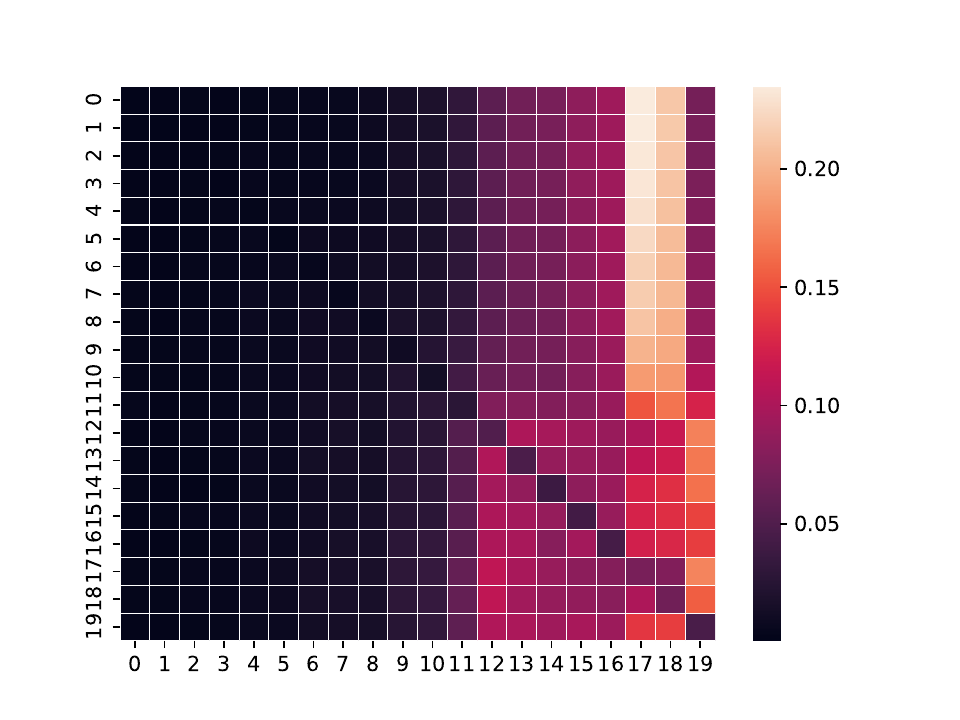}\label{fig:beauty bert}}
\vspace{-1.3em}
\caption{Attention heatmaps of the last 20 items of ML-1M and Beauty on SASRec and Bert4Rec. X-axis and Y-axis denote the item indices. The intensity of the color represents the magnitude of attention. Each entry shows the attention from X to Y. Padding tokens are inserted at the beginning for sequences shorter than 20.}
\label{fig:heatmap}
\vspace{-1.3em}
\end{figure*}

\subsection{DWRS-D}
\label{sec:dwrs-d}
\subsubsection{\textbf{Watermark Structure}}
\label{sec:wm definition}
To help us decide how to define the watermark body $x_{wm}$ and target response $y_{wm}$, we show the attention heatmaps of two datasets ML-1M and Amazon Beauty on two sequential recommendation models SASRec and Bert4Rec in \autoref{fig:heatmap}. The average sequence length is 165.5 for ML-1M, and 8.8 for Amazon Beauty. Bert4Rec is a bi-directional attention model, while SASRec is one-directional. Please refer to \autoref{sec:experiments} for more information on the datasets and models. Both datasets suggest a strong spatial dependency in the sequences, i.e., the embedding of a target item is dependent on its surrounding items.

Generally, the attention aggregation in a transformer layer is:
\begin{equation}
    \alpha = \text{softmax}(\frac{QK^{T}}{\sqrt{d_{k}}})
\end{equation}
\begin{equation}
    X = \alpha V
\end{equation}
where $\alpha\in\mathbb{R}^{n\times n}$ is the attention matrix, $V\in\mathbb{R}^{n\times d_{v}}$ is the value matrix, and $X\in\mathbb{R}^{n\times d_{v}}$ is the aggregated embedding matrix. For a random item $i$, its embedding $x_{i}$ is computed as
\begin{equation}
\label{eq:embedding i}
    x_{i}=\alpha_{i}V
\end{equation}
$x_{i}\in \mathbb{R}^{d_{v}}$ is its embedding vector, and $\alpha_{i}\in\mathbb{R}^{n}$ is its attention vector.

According to the attention heatmaps in \autoref{fig:heatmap}, the embedding of a target item is computed by aggregating its surrounding items. For any item $i$, we define its \textbf{Receptive Field} (RF) as the range where all items within it have attention values much greater than zero. The specific items within the RF are called \textbf{filler items}. We can observe that the RF always includes a set of consecutive items around item $i$. Specifically, for \autoref{fig:ml1m sas}, the RF includes the three items before item $i$, whereas for \autoref{fig:ml1m bert}, the RF includes the three items before $i$ as well as the three items after $i$. Since other items contribute little to item $i$, \autoref{eq:embedding i} can be rewritten as
\begin{equation}
\label{eq:rf}
    x_{i}\approx\sum_{j\in\text{RF}}\alpha_{ij}V_{j}
\end{equation}
where $\alpha_{ij}$ is a scalar representing the attention value of item $j$ to item $i$, and $V_{j}$ is the value vector of item $j$. Now suppose that the item $i$ is the target response $y_{wm}$ of our watermark. According to \autoref{eq:rf}, it will only aggregate information from its RF. Therefore, the watermark body $x_{wm}$ needs to be within its RF for the model to learn the watermark. Besides, we can observe that, items before $y_{wm}$ are always included in its RF, no matter it is the one-directional model or bi-directional model. This is consistent with the nature of the sequential prediction task, where former items always provide critical information on the next item. This observation provides us with a valuable insight: the watermark items in the watermark body $x_{wm}$ and the target response $y_{wm}$ should be placed consecutively for $x_{wm}$ to stay within the RF of $y_{wm}$. To this end, we define the structure of our watermark as follows:
\begin{definition}[DWRS-D watermark structure]
    We define the watermark as a set of sequentially ordered consecutive watermark items $S_{wm}=\{i^{wm}_{1},i^{wm}_{2},...,i^{wm}_{l}\}$, where $l$ denotes the watermark length. The watermark body $x_{wm}$ is defined as $\{i^{wm}_{1},i^{wm}_{2},...,i^{wm}_{l-1}\}$, while the last item $i^{wm}_{l}$ serves as the target response $y_{wm}$. 
\end{definition}

\subsubsection{\textbf{Watermark Construction}}
\label{sec:construct}
In this section, we discuss the detailed design of DWRS-D. 

\paragraph{\textbf{Inserting the Watermark}}
\label{sec:inserting ratio}
According to \autoref{fig:heatmap}, the RF is determined by both the dataset and the model. As a defender, we should assume no knowledge about the model architecture that an attacker would use, thus it is impossible for us to obtain the true RF. 

Now let us discuss how the true RF can affect the memorization of the watermark. Given a model and a dataset, suppose the true RF contains $r_{1}$ items before the target item, as well as $r_{2}$ items after the target item. The watermark length is denoted by $l$, so the length of the watermark body is $l-1$. One possible case is that the watermark covers the entire RF, for example, $r_{1}\leq l-1$ and $r_{2}=0$. In this case, all items that contribute to $y_{wm}$ are from $x_{wm}$. In this case, the watermark can be easily memorized. In other cases, for example $r_{1}>l-1$ or $r_{2}>0$, the RF contains filler items other than $x_{wm}$. They will contribute to the aggregation of $y_{wm}$, interfering with the watermark memorization. For example, a certain filler item might appear multiple times in different sequences, so the model would learn correlations between this item and $y_{wm}$, which would introduce additional bias and lead to performance degradation when evaluating the watermark using only $x_{wm}$. To alleviate this issue, we could force the filler items to follow a \textbf{uniqueness constraint}, that is, an item would appear at most once as a filler item. The complete algorithm is shown in \autoref{alg:DWRS-D}.

Specifically, we will first define a proper RF range of the specific dataset according to the attention heatmaps. For example, we can define the RF of ML-1M as five items before the target item and five items after the target item. We start with a random sequence and insert $S_{wm}$ into a random position. We record the filler items that are within the RF but do not belong to $S_{wm}$. Then in each iteration, we draw one sequence from other unselected sequences. For this sequence, we iteratively go through all positions, until we find a position such that the RF does not contain any filler item that has been recorded. We stop until the number of sequences reaches the predefined insert ratio $p$ or we cannot find any of the remaining sequences that meet the uniqueness constraint. In this way, there will be no duplicate filler items, so the noise introduced by the filler items will be alleviated.

\paragraph{\textbf{Watermark Items}} The watermark items in $S_{wm}$ can be any item of the defender's choice. However, in general, unpopular items with fewer interactions are more favorable. On the one hand, unpopular items are less observed in benign users' recommendation lists than popular items \cite{yu2023self}, which helps improve the discriminability of the watermark by achieving a low watermark success rate on oracle models. On the other hand, unpopular items receive few training signals due to the limited number of interactions they have. Therefore, they can be easily adapted to the watermark pattern. We will address this problem again in \autoref{sec:settings} and show experiment results in \autoref{appendix:popular items} for different popularity.

\paragraph{\textbf{Watermark Length}} The watermark length $l$ can also be any positive integer. But according to \autoref{fig:heatmap}, only items in the RF can contribute to the target item, so increasing the length of the watermark would have limited performance gain on the watermark success rate if the watermark items fall out of the RF of $y_{wm}$. In addition, a long watermark would be easily detected by statistic methods such as frequent pattern mining \cite{han2001prefixspan,ayres2002sequential}. Therefore, a short watermark is more favorable in achieving high watermark utility and good unnoticeability.

\subsubsection{\textbf{Watermark Evaluation}}
When the dataset owner suspects that a target model is trained on their watermarked dataset, they can claim the ownership by querying the model with the watermark. Specifically, we assume that the dataset owner only has black-box access to the target model, i.e., they can only query the model and observe the output. Thus, to evaluate the watermark, the model owner can randomly select some benign user sequences and append $x_{wm}=\{i^{wm}_{1},i^{wm}_{2},...,i^{wm}_{l-1}\}$ to the end of those benign sequences, and then query the target model. The goal is to check the ranking of the response item $y_{wm}=i^{wm}_{l}$ in the output ranking lists. If $i^{wm}_{l}$ is ranked at the top, the model owner can safely claim the ownership.

\section{User Watermarking}
\label{sec:user watermarking}
So far, we have explored the dataset watermarking to protect the copyright of the entire dataset. We are also interested in the problem of user watermarking, where an individual user may want to protect his/her own data by embedding a watermark in it. This is a very challenging problem. The quality of the watermark can be influenced by multiple factors such as the interaction history of the target user, the distribution of the entire dataset, as well as the implementation of the model. As a pioneer work, we define the problem of user watermarking and present our initial solution to it. More powerful user watermarking techniques are a promising future direction.

\subsection{Problem Definition}
We define the problem of user watermarking as to inject a watermark into a target user's interaction sequence, so that a model trained on the dataset that involves this watermarked user would have a high watermark success rate. We assume that the target user does not have access to other users' interactions, i.e., the target user cannot view or edit other users' interactions. The target user is only capable of determining his or her own interaction sequence by interacting with any item at any predetermined time slot. We also assume that the target user does not have any knowledge regarding the recommender system. However, we assume that the target user has access to the item popularity, e.g., how many times an item has been interacted by all users. This is a reasonable assumption on most e-commerce and social platforms (e.g., a user can see how many times an item has been purchased). The target user can leverage this information to help design the watermark.

The user watermark should also meet the universality, unnoticeability, and discriminability requirements mentioned in \autoref{sec:def 1}. We do not consider robustness in this work, as we assume that the server would honestly train their model with the raw dataset. Generally, things become more challenging than dataset watermarking as we are only permitted to inject the watermark into one sequence with limited prior knowledge. Next, we present our analysis and solution to this problem.

\subsection{DWRS-U}
We define the user watermark in a similar way as the dataset watermark. The user watermark is a sequence of watermark items $S_{wm}=\{i^{wm}_{1},i^{wm}_{2},...,i^{wm}_{l}\}$ with length $l$. $S_{wm}$ would only be inserted into the target user's interaction sequence. 

\subsubsection{\textbf{Watermark Item Selection}}
\label{sec:construct 2}
Since the watermark signal strength is largely reduced compared with dataset watermark as it only appears once in the entire dataset, we need to compensate for the strength loss by other means, e.g., choices of watermark items, the insert position, as well as the watermark length.

A detailed description of our DWRS-U algorithm for user watermarking is shown in \autoref{alg:DWRS-U}. Since the watermark only appears once in the entire dataset, it is difficult for the model to memorize it. To alleviate this issue, the watermark items should ideally not appear in other sequences, thereby facilitating the memorization of the watermark pattern. Therefore, we select the most unpopular items as watermark items. Specifically, if the most unpopular item has three interactions, then all those items with three interactions are considered as candidate watermark items. If the number of items with three interactions are smaller than the watermark length $l$, we can expand the candidate item pool to include those items with four interactions. Since users have access to the item popularity, this strategy is feasible to implement.

\begin{algorithm}[h]
\caption{DWRS-U}
\label{alg:DWRS-U}
\begin{algorithmic}[1]
\REQUIRE
Target user $u_{t}$, interaction sequence $S_{t}$, item set $\mathcal{I}$, watermark length $l$, unpopular subsequence length $n$;
\ENSURE
Watermarked target sequence $S_{t}'$;
\STATE Rank all items from $\mathcal{I}$ according to their popularity;
\STATE Sample $l$ items from the most unpopular items \\to form watermark sequence $S_{wm}$;
\STATE $S_{t}'\gets S_{t}$;
\WHILE{$S_{wm}$ exists in $S_{t}'$}
    \STATE Resample $S_{wm}$;
\ENDWHILE
\STATE $\mathcal{S}_{sub} \gets$ the set of all subsequences with length $n$ from $S_{t}'$;
\FOR{$S\in \mathcal{S}_{sub}$}
\STATE Compute the average item popularity of $S$;
\ENDFOR
\STATE Rank all subsequences in $\mathcal{S}_{sub}$ according to the average \\item popularity;
\STATE $\mathcal{C}_{n}\gets$ the most unpopular subsequence;
\STATE Insert $S_{wm}$ into $S_{t}'$ right before $\mathcal{C}_{n}$;
\end{algorithmic}
\end{algorithm}

\subsubsection{\textbf{Watermark Length \& Insert Position}}
A larger watermark length $l$ naturally enhances memorization, as it provides more dependencies suggested by \autoref{fig:heatmap}. This would make the prediction of $i^{wm}_{l}$ easier to overfit given $\{i^{wm}_{1},i^{wm}_{2},...,i^{wm}_{l-1}\}$. In fact, we have empirically verified that, under the default setting of our experiment, by setting $l>30$ and using unpopular items as watermark items, the watermark is almost guaranteed to be successful with Recall@10 $>$ 0.9, regardless of the insert position. It suggests that we can always create a strong watermark by increasing the watermark length $l$. However, a long watermark would fail the requirement of unnoticeability and may increase the risk of being detected. Therefore, we next discuss the possibility of choosing a dedicated insert position which can help alleviate the need for a long watermark.

If we want to reduce the watermark length $l$, we need to compensate for the dependency loss by other means. Since we are evaluating the watermark by predicting $y_{wm}=i^{wm}_{l}$ given $x_{wm}=\{i^{wm}_{1},i^{wm}_{2},...,i^{wm}_{l-1}\}$, let us now focus on the last watermark item $y_{wm}=i^{wm}_{l}$. \autoref{fig:heatmap} suggests that items before $i^{wm}_{l}$ contribute significantly to its embedding. However, note that $i^{wm}_{l}$ also contributes to the embeddings of items after it. For convenience, we name a sequence of those items after $i^{wm}_{l}$ as $\mathcal{C}_{n}$, where $n$ denotes the number of items considered. Since the embeddings of items in $\mathcal{C}_{n}$ are obtained by aggregating information from previous items, learning to predict items in $\mathcal{C}_{n}$ in the training process would require updating the embedding of $i^{wm}_{l}$ as well. Therefore, $i^{wm}_{l}$ and $\mathcal{C}_{n}$ would also establish strong correlations, and we may compensate for the dependency loss of reducing $l$ by utilizing $\mathcal{C}_{n}$ to increase the dependency capacity. To achieve this, items in $\mathcal{C}_{n}$ should also be unpopular, as their embeddings are less robust and can be easily fused with the watermark information, thereby increasing the dependency capacity of $S_{wm}$. To this end, we compute the average item popularity of each subsequence, and name the one with the least average popularity as the most \textbf{unpopular} subsequence of the target sequence. We insert the watermark right before the most unpopular subsequence. For example, if given $n$=10, we would first iterate through the target user sequence and extract all subsequences with length 10. Then for each subsequence, we compute the average popularity (number of interactions) of items in it, and select the most unpopular subsequence as $\mathcal{C}_{n}$. The watermark would be inserted right before $\mathcal{C}_{n}$. In this way, we may shorten the length $l$ of the watermark $S_{wm}$. As $l$ decreases, the dependency capacity of $S_{wm}$ also decreases. Thus the model would try to compensate for the capacity loss by utilizing items in $\mathcal{C}_{n}$. When evaluating the watermark, we would discard $\mathcal{C}_{n}$ and append $\{i^{wm}_{1},i^{wm}_{2},...,i^{wm}_{l-1}\}$ to the query sequence and evaluate the prediction of $i^{wm}_{l}$. 

\subsubsection{\textbf{Discussions}}
Some considerations may be raised regarding the above method. On the one hand, it seems that we may manually insert some unpopular items after $i^{wm}_{l}$ to serve as $\mathcal{C}_{n}$. However, we argue that this does not achieve the goal of reducing the watermark length. Although $S_{wm}$ itself is indeed shortened, we still have to insert $n$ more items to form $\mathcal{C}_{n}$, and the total number of inserted items would become $l+n$. On the other hand, it seems that inserting $S_{wm}$ after the most unpopular subsequence can achieve the same effect of increasing the dependency capacity. But putting unpopular item before the watermark is actually equivalent to expanding the watermark with those items. One still has to include those items in the query when evaluating the watermark in order to achieve a high watermark validity. We are also aware of a weakness of this method. Specifically, if the target user does not have a subsequence filled with unpopular items, or if those items are not unpopular enough, this method may fail and we may have to increase the length of the watermark. 

It seems that we could also use DWRS-U for the dataset watermarking. But as we will see in our experiment results, DWRS-D already achieves a very good watermarking quality (high watermark validity and model utility) with a shorter watermark length and a smaller insert ratio. In this case, using DWRS-U to do dataset watermarking would not bring a significant improvement on the watermark validity, but would instead increase the chance of exposure, because the sampling strategies in DWRS-U can introduce potential bias that could be detected. Therefore, DWRS-D is generally preferred for dataset watermarking. 

\begin{table}[ht]
\caption{Dataset statistics.}
\vspace{-1.5em}
\begin{center}
 \begin{tabular}{|c|c|c|c|c|c|} 
 \hline
 Dataset & \# users & \# items & avg. len. & max len. & density \\
 \hline
 ML-1M & 6,040 & 3,416 & 165.5 & {\cheng 2,277} & 4.84\% \\
 \hline
 Beauty & 40,226 & 54,542 & 8.8 & 293 & 0.02\% \\
 \hline
 Steam & 334,542 & 13,046 & 12.6 & {\cheng 2,045} & 0.10\% \\
 \hline
\end{tabular}
\label{table:datasets}
\end{center}
\vspace{-1.5em}
\end{table}

\begin{figure*}[t]
\centering
    \includegraphics[width=1.0\textwidth]{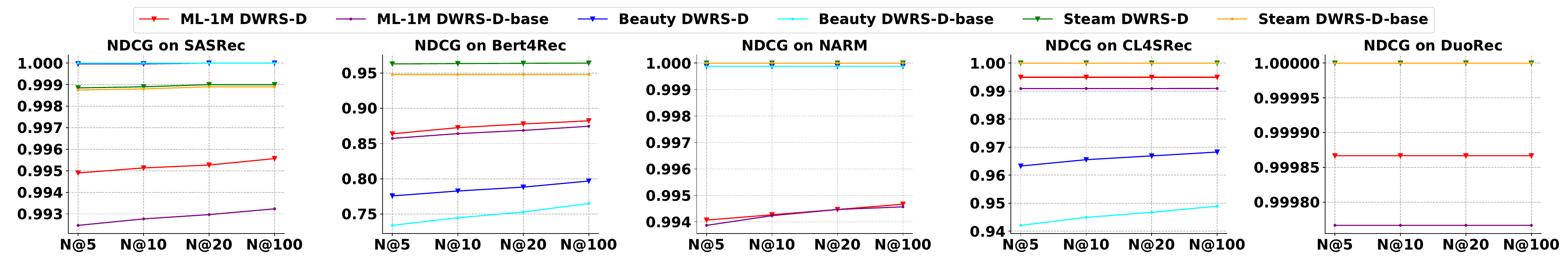}
\vspace{-2.5em}
\caption{DWRS-D: watermark validity (NDCG).}
\label{fig:main ndcg}
\vspace{-1.5em}
\end{figure*}

\section{Experiments}
\label{sec:experiments}
We conducted experiments to assess the performance of DWRS-D and DWRS-U across five sequential recommender systems and three datasets. Our study aims to address the following research questions:

\begin{itemize}
    \item \textbf{RQ1}: How does DWRS-D compare to baseline methods in terms of watermark validity and model utility?
    \item \textbf{RQ2}: What impact do watermark length $l$ and insert ratio $p$ have on DWRS-D?
    \item \textbf{RQ3}: How resilient is DWRS-D against watermark removal attacks?
    \item \textbf{RQ4}: How does DWRS-U perform in terms of watermark validity and model utility?
    \item \textbf{RQ5}: How do watermark length $l$ and insert position influence DWRS-U?
\end{itemize}

\subsection{Datasets}
We conducted experiments on three datasets, namely MovieLens-1M\footnote{\url{https://grouplens.org/datasets/movielens/}}, Amazon Beauty\footnote{\url{https://cseweb.ucsd.edu/~jmcauley/datasets/amazon/links.html}}, and Steam\footnote{\url{https://cseweb.ucsd.edu/~jmcauley/datasets.html\#steam_data}}. We follow {\cheng a} previous work \cite{yue2021black} to preprocess the datasets. The statistics are shown in \autoref{table:datasets}. Each dataset contains a large number of users, and each user is associated with one interaction sequence.

\begin{table}[ht]
\caption{Average watermark utility (NDCG) on all oracle models.}
\vspace{-2em}
\begin{center}
 \begin{tabular}{|c|c|c|c|c|c|c|c|c|c|c|c|} 
 \hline
 Dataset & N@5 & N@10 & N@20 & N@100 \\
 \hline
   ML-1M & 0.00\% & 0.00\% & 0.01\% & 0.04\% \\
   \hline
 Beauty & 0.00\% & 0.00\% & 0.00\% & 0.02\% \\
 \hline
 Steam & 0.00\% & 0.00\% & 0.00\% & 0.00\% \\
 \hline
\end{tabular}
\label{table: watermark utility on oracle}
\end{center}
\vspace{-2em}
\end{table}

\subsection{Models and Baselines}
We evaluate DWRS on five representative sequential recommendation models, SASRec \cite{kang2018self}, Bert4Rec \cite{sun2019bert4rec}, NARM \cite{li2017neural}, CL4SRec \cite{xie2022contrastive}, and DuoRec \cite{qiu2022contrastive}. Among them, NARM is an RNN-based encoder-decoder framework, and the other four models are transformer-based models. SASRec adopts one-directional attention, while Bert4Rec adopts masking mechanism and bi-directional attention. CL4SRec incorporates sequence-level contrastive learning, while DuoRec utilizes model-level contrastive learning.

Since existing data watermarking techniques cannot be adapted to recommendation, we compare DWRS-D with \textbf{DWRS-D-base}, where the watermark is inserted into random positions of random sequences. We compare DWRS-U with \textbf{other insert positions}.

\subsection{Evaluation Metrics}
We use the traditional leave-one-out protocol, where the last item of each user sequence is used as the test item and the penultimate item is used as the validation item. The model utility is evaluated by Recall@k and NDCG@k on the test item. Both metrics evaluate how the target item is ranked in the top-k ranking list. The largest possible value for Recall and NDCG is 1.0, indicating that the target item is ranked at the top. To evaluate the validity of the watermark, we append the watermark body $x_{wm}=\{i^{wm}_{1},i^{wm}_{2},...,i^{wm}_{l-1}\}$ to benign sequences and query the model to obtain the prediction for the target response item $y_{wm}=i^{wm}_{l}$. We compute the average Recall@k and NDCG@k for $i^{wm}_{l}$ across all user sequences.

\subsection{RQ1: Evaluating DWRS-D}
\subsubsection{\textbf{Settings}}
\label{sec:settings}
We follow previous studies \cite{yue2021black,ren2024sslrec} and use the suggested hyperparameter settings for each model and each dataset. For the watermark $S_{wm}$, we set by default its length $l$=3. The watermark items in $S_{wm}$ are randomly selected from the top-10\% unpopular items (using popular items as watermark leads to poor performance as shown in \autoref{appendix:popular items}), but with a constraint that $S_{wm}$ does not appear in the original dataset, i.e., $S_{wm}$ is a completely new pattern that does not exist in the original dataset. Since some real-world datasets do not allow repeated items in one sequence, we insert $S_{wm}$ into users that have not interacted with any of the watermark items. We set the receptive field of ML-1M as five items before $S_{wm}$ as well as five items after $S_{wm}$, and that of Beauty and Steam as the entire sequence. We set the insert ratio as $p$=1\%, a widely accepted threshold in the literature on adversarial attacks for recommender systems \cite{tang2020revisiting,yue2021black}, i.e., we iteratively select new users to insert the watermark, until we either watermarked 1\% users or cannot find any of the remaining users that meet the uniqueness constraint. For a fair comparison, once the watermark $S_{wm}$ is generated in each run, it is fixed and used across all models. We report average results of five runs.

\begin{figure}[ht]
\vspace{-1.5em}
\centering
     \subfloat[]{\includegraphics[width=0.5\linewidth]{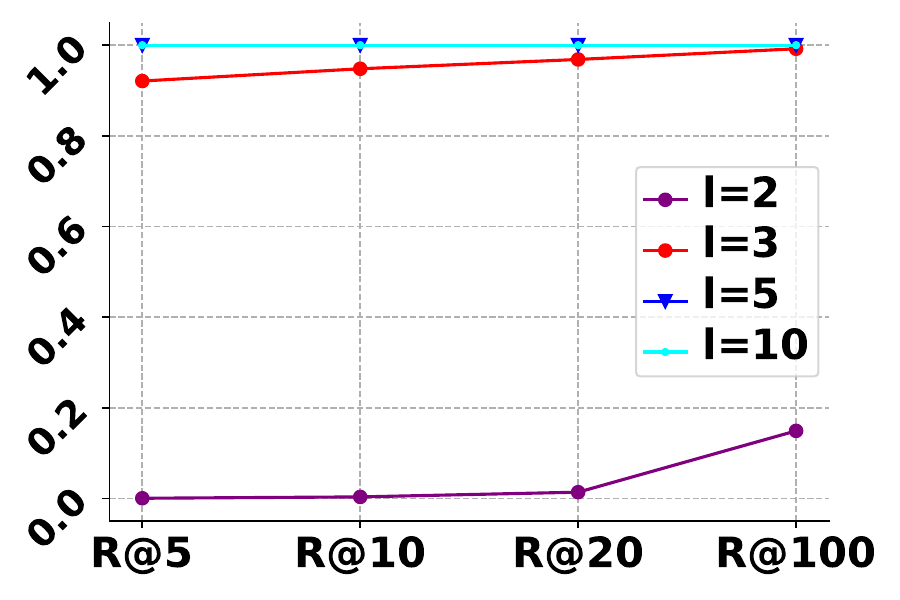}\label{fig:plot l}}
	\subfloat[]{\includegraphics[width=0.5\linewidth]{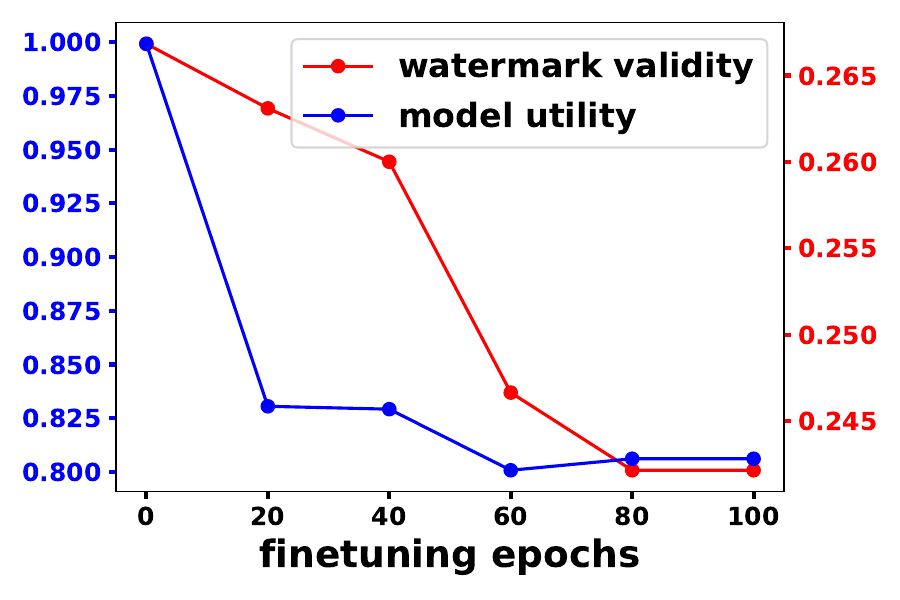}\label{fig:finetune}}
\vspace{-1.3em}
\caption{DWRS-D: (a) watermark validity (Recall) of Bert4Rec on ML-1M under different watermark length $l$. (b) watermark validity and model utility (Recall@10) of SASRec on ML-1M after finetuning.}
\label{fig:different l and finetuning}
\vspace{-1.5em}
\end{figure}

\begin{figure*}[t]
\centering
    \includegraphics[width=1.0\textwidth]{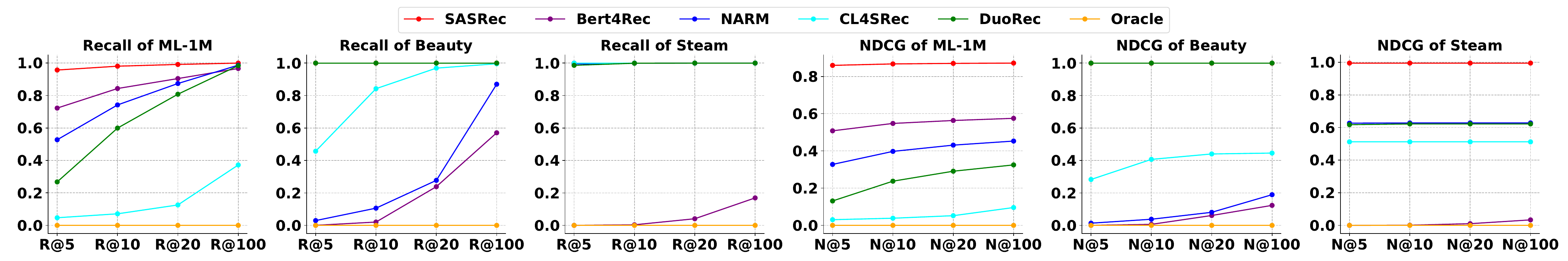}
\vspace{-2.5em}
\caption{DWRS-U: watermark validity. For ML-1M and Steam, we set $l$=10 and $n$=10. For Beauty, we set $l$=20 and $n$=10. Oracle denotes testing the watermark on the oracle model.}
\label{fig:user wm validity}
\vspace{-1.5em}
\end{figure*}

\subsubsection{\textbf{Results}}
We first show the watermark validity (NDCG) in \autoref{fig:main ndcg}. The results of Recall are in \autoref{appendix: DWRS-D watermark validity recall} \autoref{fig:main recall}, as they behave similarly with NDCG. We can observe that both DWRS-D and DWRS-D-base achieve high NDCGs across all models and datasets (>0.7 for Bert4Rec and >0.9 for other models). DWRS-D outperforms DWRS-D-base because its NDCGs are consistently equal to or better than that of DWRS-D-base. In addition, according to the discriminability requirement, a good watermark should not exist in oracle models. For each dataset, we compute the average watermark validity on all oracle models. The results are shown in \autoref{table: watermark utility on oracle}. The NDCGs are close to zero, indicating that we cannot detect the watermark from the oracle models, which satisfies the discriminability requirement.

Besides, we show the model utility of all datasets and models in \autoref{appendix: model utility} \autoref{table:DWRS-D all model utility}. We show both the utility of the oracle models which are trained on clean datasets and the utility of the watermarked models which are trained on the watermarked datasets by DWRS-D and DWRS-U respectively. For now, let us focus on Oracle and DWRS-D. We can observe a comparable performance between the oracle model and the watermarked model with no significant gap, demonstrating the effectiveness of DWRS-D in preserving the model utility and achieving a good unnoticeability.

\subsection{RQ2: Hyperparameter Studies for DWRS-D}
\subsubsection{\textbf{Studies on watermark length $l$}}
We take ML-1M on Bert4Rec as an example as shown in \autoref{fig:plot l}. It is clear that as $l$ increases, the watermark validity also increases, suggesting that increasing the watermark length can boost the validity of the watermark. An interesting observation is that, $l=2$ performs poorly with Recall close to 0, which suggests that using a single item as the watermark body is insufficient in memorizing the watermark. Additionally, $l=3$ achieves a very high Recall around 0.9, and further increasing $l$ brings minor improvement. Thus, we select $l=3$ as our default watermark length to enhance the unnoticeability.

\begin{table}[ht]
\caption{Recall@10 in percentage (\%) of ML-1M on SASRec with different insert ratio $p$.}
\vspace{-1.5em}
\begin{tabular}{|c|c|c|c|c|c|}
\hline
p & 0.1\%  & 0.5\% & 1\% & 1.5\% & 2\% \\
\hline
Recall@10  & 85.77  & 99.66 & 99.74 & 100 & 100 \\
\hline
\end{tabular}
\label{table:insert ratio p}
\end{table}

\subsubsection{\textbf{Studies on insert ratio p}}
Taking ML-1M on Bert4Rec for an example, we show how the insert ratio $p$ affect the watermark validity in \autoref{table:insert ratio p}. The validity increases as $p$ increases. When $p$ is small, increasing $p$ could significantly improve the watermark validity. But the performance gain becomes less as $p$ keeps increasing. If $p$ is too small, the watermark validity would be low; if $p$ is too large, the unnoticeability would be affected. In practice, a proper $p$ should be carefully selected. We choose 1\% as our default setting, as it is a widely accepted threshold in the literature on adversarial attacks for recommender systems \cite{tang2020revisiting,yue2021black}.

\subsection{RQ3: DWRS-D Against Attacks}
\label{sec:against removal attacks}
We evaluate how DWRS-D performs against three types of attacks: Finetuning, Distillation, and Sequential Rule Mining. The results of finetuning are shown in this section, while distillation and sequential rule mining are shown in \autoref{appendix:against attacks}. According to our experiments, DWRS-D is resistant to all three attacks.

\subsubsection{\textbf{Against Finetuning}}
\label{sec:finetuning}
We use 80\% data with inserted watermarks to train a watermarked model. We then finetune the model with the rest 20\% clean data. We show the Recalls of SASRec on ML-1M in \autoref{fig:finetune}. Other models and datasets perform similarly. Both the watermark validity and the model utility are displayed on the figure. As the number of epochs increases, the watermark validity and model utility both decrease. Watermark validity drops from 1.0 to 0.8 after 100 finetuning epochs. However, it can still provide strong evidence for data theft, because an oracle model should have zero Recall as shown in \autoref{table: watermark utility on oracle}. However, the model utility drops from 0.265 to below 0.245, which is usually an intolerable utility loss since the recommender systems are highly commercialized products. Such performance drop may bring huge commercial loss. Therefore, DWRS-D is effective against finetuning attacks.

\subsection{RQ4: Evaluating DWRS-U}
\subsubsection{\textbf{Settings}}
We keep the hyperparameter settings unchanged for each model and each dataset as when evaluating DWRS-D. We use the longest user sequence as the target user, a very challenging scenario where the overwhelming number of benign items may introduce a large amount of noise when memorizing the watermark. For the watermark $S_{wm}$, we randomly select watermark items from those with the least number of interactions as introduced in \autoref{sec:construct 2}. We also ensure that all the watermark items do not exist in the target user sequence. By default, we set the watermark length $l$=10 for ML-1M and Steam, and 20 for Beauty. The insert position is computed with $n$=10, which means that we compute the most unpopular subsequence with length 10. We empirically found that $n$=10 is a good choice for all tested cases, and varying $n$ around 10 does not significantly impact performance.

\begin{figure}[ht]
\centering
     \subfloat[]{\includegraphics[width=0.45\linewidth]{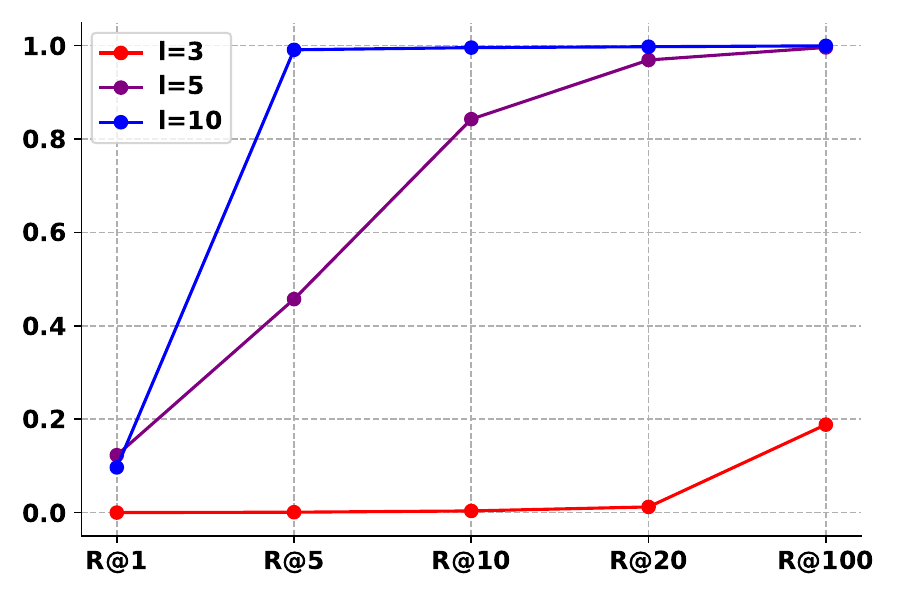}\label{fig:DWRS-U l}}
	\subfloat[]{\includegraphics[width=0.5\linewidth]{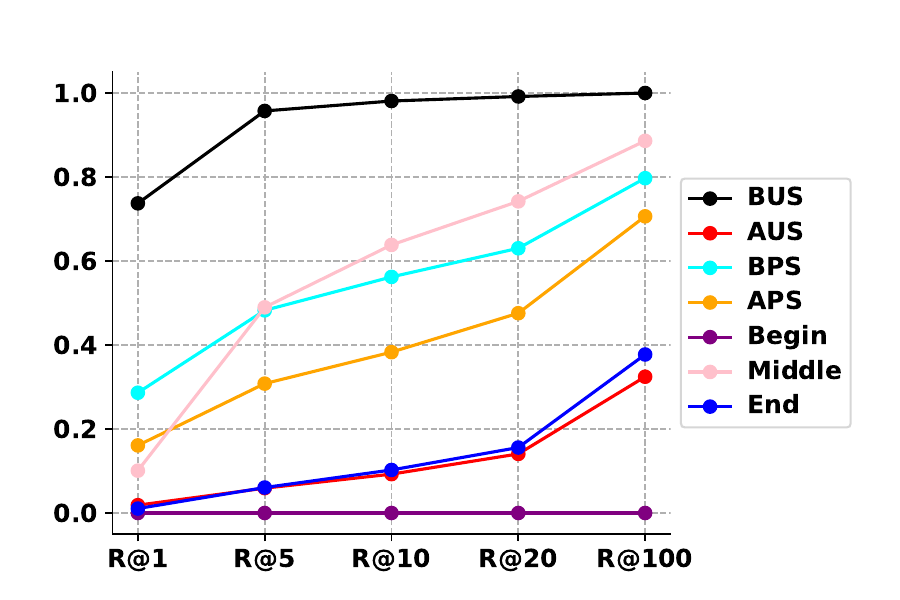}\label{fig:inserting position}}
\vspace{-1.3em}
\caption{DWRS-U: (a) watermark validity (Recall) of CL4SRec on Beauty with $n$=10 under different watermark length $l$. (b) Recalls for different insert positions of SASRec on ML-1M. BUS: before the most unpopular subsequence; AUS: after the most unpopular subsequence; BPS: before the most popular subsequence; APS: after the most popular subsequence; Begin: at the beginning of the sequence; Middle: at the middle of the sequence; End: at the end of the sequence.}
\label{fig:dwrsu_l and insert positions}
\end{figure}

\subsubsection{\textbf{Results}}
The watermark validity is shown in \autoref{fig:user wm validity}. We can observe that DWRS-U has a good watermark validity in most cases. However, the watermark validity is generally not as strong as that of DWRS-D, but it remains effective because the Recall and NDCG on the oracle model are close to zero. Therefore, we can still safely claim the ownership due to the significant margin. Besides, we can observe a large performance gap on different models. For example, Bert4Rec has the worst watermark validity on Beauty and Steam among all models. This performance gap might be due to the specific model design and its convergence status. Specifically, Bert4Rec is known to be difficult to train and converge \cite{petrov2022systematic}. If the model itself is underfitted, it is also difficult for it to overfit and memorize the watermark. But this problem can be alleviated by increasing the watermark length $l$, as we will see in \autoref{sec: DWRS-U hyper study}.

The model utility is shown in \autoref{appendix: model utility} \autoref{table:DWRS-D all model utility}. From the performance of Oracle and DWRS-U, we can see that DWRS-U also has a comparable performance with the oracle model, demonstrating the ability to preserve the model utility.

\subsection{RQ5: Hyperparameter Studies for DWRS-U}
\label{sec: DWRS-U hyper study}
We present the hyperparameter studies on the watermark length $l$ and the insert position for DWRS-U. For $l$, we take the results of CL4SRec on Beauty for an example, as shown in \autoref{fig:DWRS-U l}. We can observe that as $l$ increases, the watermark validity also increases. This suggests that increasing $l$ can help enhance the watermark validity. For the insert position, we take the results of SASRec on ML-1M for an example, as shown in \autoref{fig:inserting position}. For simplicity and consistency, we show the performance of several representative insert positions. All subsequences are computed with length 10. We can observe that BUS significantly outperforms other insert positions, demonstrating the superiority of inserting the watermark right before the most unpopular subsequence.

\section{Conclusion}\label{sec:con}
We present comprehensive definitions and analysis on the problems of dataset watermarking and user watermarking for sequential recommender systems. We propose a data watermarking method named Data Watermarking for Recommender Systems (DWRS) and its two variants, DWRS-D and DWRS-U for dataset watermarking and user watermarking respectively. Extensive experiments demonstrate the superior performance of DWRS in protecting the data ownership while preserving the model utility, as well as its robustness against removal attacks.

\begin{acks}
This research is supported by the Ministry of Education, Singapore, under its Academic Research Fund (Tier 2 Award MOE-T2EP20221-0013 and Tier 1 Award (RG20/24)). Any opinions, findings and conclusions or recommendations expressed in this material are those of the author(s) and do not reflect the views of the Ministry of Education, Singapore. The Australian Research Council partially supports this work under the streams of Future Fellowship (Grant No. FT210100624), the Discovery Project (Grant No. DP240101108), and the Linkage Projects (Grant No. LP230200892 and LP240200546).
\end{acks}

\balance
\bibliographystyle{ACM-Reference-Format}
\bibliography{ref.bib}

\appendix

\begin{table}[ht]
\caption{Recall in percentage (\%) of ML-1M on SASRec before and after distillation}
\vspace{-1.5em}
\begin{center}
\begin{tabular}{|c|c|c|c|}
\hline
 & Recall & before & after \\
 \hline
\multirow{4}{*}{watermark validity} & R@5 & 99.68 & 19.67 \\
\cline{2-4}
 & R@10 & 99.74 & 29.47 \\
\cline{2-4}
 & R@20 & 99.81 & 40.85 \\
\cline{2-4}
 & R@100 & 99.97 & 71.37 \\
 \hline
\multirow{4}{*}{model utility} & R@5 & 19.09 & 14.09 \\
\cline{2-4}
 & R@10 & 27.04 & 22.12 \\
\cline{2-4}
 & R@20 & 37.92 & 32.78 \\
\cline{2-4}
 & R@100 & 63.8 & 61.27 \\
 \hline
\end{tabular}
\label{table:distillation}
\end{center}
\end{table}

\begin{table*}[t]
\caption{Performance in percentage (\%) of popular items vs. unpopular items as watermark (ML-1M on Bert4Rec).}
\vspace{-1em}
\begin{tabular}{|c|c|c|c|c|c|c|c|c|c|c|c|c|}
\hline
 & \multicolumn{4}{c|}{watermark validity} & \multicolumn{4}{c|}{model utility} & \multicolumn{4}{c|}{oracle model validity} \\
 \hline
 & R@10 & R@20 & N@10 & N@20 & R@10 & R@20 & N@10 & N@20 & R@10 & R@20 & N@10 & N@20 \\
 \hline
unpopular & 94.79 & 96.85 & 87.25 & 87.77 & 19.82 & 30.34 & 10.42 & 13.06 & 0.00 & 0.00 & 0.00 & 0.0 \\
\hline
popular & 20.55 & 28.55 & 11.92 & 13.94 & 19.70 & 30.41 & 10.44 & 13.13 & 6.26 & 10.84 & 2.88 & 4.03 \\
\hline
\end{tabular}
\label{table:popular items as watermark}
\end{table*}

\begin{figure*}[t]
\centering
    \includegraphics[width=1.0\textwidth]{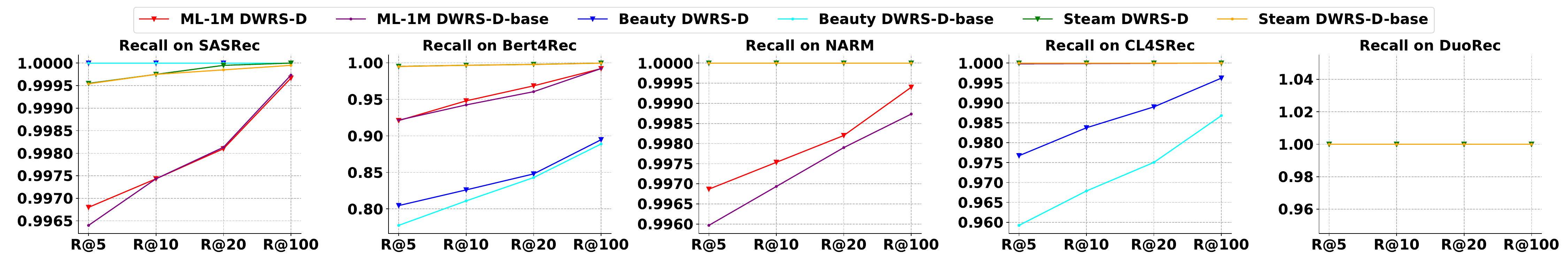}
\vspace{-2.5em}
\caption{DWRS-D: watermark validity (Recall).}
\label{fig:main recall}
\vspace{-1.5em}
\end{figure*}

\section{Popular items vs. unpopular items}
\label{appendix:popular items}
We show the model performance of popular items vs. unpopular items as watermark in \autoref{table:popular items as watermark}. We randomly select popular items from top-10\% popular ones, and unpopular items from top-10\% unpopular ones. We take ML-1M on Bert4Rec as an example. Other datasets and models perform similarly. It is clear that while the model utility is comparable, unpopular items show significantly better watermark validity than popular items. Besides, when testing the watermark on oracle models (as shown in the oracle model validity column), popular items show a non-negligible chance of being detected, thus affecting the discriminability of the watermark. Instead, using unpopular items as watermark leads to zero Recall and NDCG on oracle models. Therefore, unpopular items are preferred as watermark items.

\section{Watermark validity (Recall) of DWRS-D}
\label{appendix: DWRS-D watermark validity recall}

The watermark validity (Recall) of DWRS-D on all datasets and models are presented in \autoref{fig:main recall}.

\section{Model utility}
\label{appendix: model utility}
The model utilities of DWRS-D, DWRS-U and Oracle on all models and datasets are shown in \autoref{table:DWRS-D all model utility}.

\section{RQ3: DWRS-D Against Attacks}
\label{appendix:against attacks}

\subsubsection{\textbf{Against Distillation}}
\label{sec:distillation}
Model distillation \cite{yue2021black} is a powerful method to remove watermarks. We test how DWRS-D performs against a model distillation method proposed by Yue et al. \cite{yue2021black}. Given a target watermarked model, we follow the default settings to generate 3,000 sequences by querying the target model, and then train a surrogate model using the generated sequences. We report the watermark validity and the model utility before and after distillation. The results of ML-1M on SASRec is shown in \autoref{table:distillation}. Other datasets and models perform similarly. Both the watermark validity and the model utility decrease significantly after distillation. The conclusion is similar to that of finetuning: 1. The non-zero Recall can still provide evidence for data theft; 2. The utility loss is fatal for commercial recommender systems. Therefore, DWRS-D is effective against model distillation.

\subsubsection{\textbf{Against Sequential Rule Mining}}
\label{sec:sequential rule mining}
Sequential rule mining \cite{fournier2014erminer} is an effective attack method in detecting the watermark if the watermark items are unpopular. Specifically, sequential rule mining discovers all rules in the form of $X\rightarrow Y$ in the dataset. The rule indicates that $Y$ appears after $X$, where $X$ and $Y$ are both sets of unordered items. For example, given $S_{wm}=\{i^{wm}_{1},i^{wm}_{2},i^{wm}_{3}\}$, we can extract five rules: $\{i^{wm}_{1}\}\rightarrow \{i^{wm}_{2}\}$, $\{i^{wm}_{1}\}\rightarrow \{i^{wm}_{3}\}$, $\{i^{wm}_{2}\}\rightarrow \{i^{wm}_{3}\}$, $\{i^{wm}_{1}\}\rightarrow \{i^{wm}_{2}, i^{wm}_{3}\}$, $\{i^{wm}_{1}, i^{wm}_{2}\}\rightarrow \{i^{wm}_{3}\}$. Each rule is associated with a confidence score, indicating the probability that $Y$ occurs, given that $X$ is observed. If using the vanilla DWRS-D, the confidence score of the above rules would be close to 100\%, thus being easily detected. To avoid it, we can make two simple changes to the algorithm: first, insert more watermarks with the same $x_{wm}$ but different $y_{wm}$; second, insert additional shuffled $x_{wm}$ (without $y_{wm}$) into the dataset. In this way, the confidence score of all rules within $S_{wm}$ would be significantly reduced. In our experiments, we insert two watermarks with the same watermark body $x_{wm}$ but different watermark response $y_{1}$ and $y_{2}$. In this case, the confidence score of $x_{wm}\rightarrow y_{wm}$ reduces from 100\% to 50\%. We also insert shuffled $x_{wm}$ to reduce the confidence scores of item pairs within $x_{wm}$. We report Recall@10 of ML-1M on all five models in \autoref{table: against sequential rule mining}. Other datasets perform similarly. From the table, both DWRS-D-base and DWRS-D achieve high Recalls. DWRS-D consistently outperforms DWRS-D-base across all models. Therefore, inserting multiple watermarks can effectively bypass sequential rule mining, while keeping the watermark validity sufficiently high.

\begin{table}[ht]
\caption{Recall@10 in percentage (\%) of ML-1M against sequential rule mining}
\vspace{-1.5em}
\begin{tabular}{|c|c|c|}
\hline
& DWRS-D-base & DWRS-D \\
\hline
SASRec & 99.95 & 99.99 \\
\hline
Bert4Rec & 91.28 & 96.17 \\
\hline
NARM & 98.90 & 99.56 \\
\hline
CL4SRec & 99.74 & 99.75 \\
\hline
DuoRec & 99.98 & 100\\
\hline
\end{tabular}
\label{table: against sequential rule mining}
\end{table}

\begin{table*}[ht]
\caption{Model utility in percentage (\%). For DWRS-D, we set $l$=3. For DWRS-U, we set $l$=10 and $n$=10 for ML-1M and Steam, and set $l$=20 and $n$=10 for Beauty. The boldfaced ones are the best.}
\vspace{-1.5em}
\begin{center}
 \begin{tabular}{|c|c|c|c|c|c|c|c|c|c|c|c|c|} 
 \hline
 Model & Dataset & Model & R@1 & R@5 & R@10 & R@20 & R@100 & N@1 & N@5 & N@10 & N@20 & N@100 \\
 \hline
 \multirow{9}{*}{SASRec} &\multirow{3}{*}{ML-1M} & Oracle & 7.11 & \textbf{19.24} & \textbf{27.45} & 37.82 & 63.83 & 7.11 & \textbf{13.29} & \textbf{15.92} & 18.54 & \textbf{23.33} \\
  && DWRS-D & 7.06 & 19.09 & 27.04 & 37.92 & 63.80 & 7.06 & 13.20 & 15.77 & 18.52 & 23.27 \\
  && DWRS-U & \textbf{7.12} & 19.02 & \textbf{27.45} & \textbf{38.05} & \textbf{64.00} & \textbf{7.12} & 13.19 & 15.89 & \textbf{18.57} & \textbf{23.33} \\
 \cline{2-13}
 &\multirow{3}{*}{Beauty} & Oracle & 0.60 & 2.28 & \textbf{3.87} & \textbf{5.95} & \textbf{12.87} & 0.60 & \textbf{1.44} & \textbf{1.95} & \textbf{2.47} & \textbf{3.72} \\
 & & DWRS-D & \textbf{0.62} & \textbf{2.30} & 3.82 & 5.83 & 12.71 & \textbf{0.62} & \textbf{1.44} & 1.94 & 2.44 & 3.68 \\
 & & DWRS-U & 0.61 & 2.12 & 3.62 & 5.63 & 12.36 & 0.61 & 1.36 & 1.84 & 2.34 & 3.56 \\
 \cline{2-13}
 &\multirow{3}{*}{Steam} & Oracle & 12.23 & 16.73 & 20.22 & 25.15 & 43.60 & 12.23 & 14.50 & 15.63 & 16.87 & 20.18 \\
  & & DWRS-D & \textbf{12.24} & 16.68 & 20.19 & 25.14 & 43.56 & \textbf{12.24} & 14.48 & 15.61 & 16.86 & 20.17 \\
  & & DWRS-U & 12.23 & \textbf{16.75} & \textbf{20.26} & \textbf{25.20} & \textbf{43.69} & 12.23 & \textbf{14.52} & \textbf{15.65} & \textbf{16.89} & \textbf{20.21} \\
 \hline
 \multirow{9}{*}{Bert4Rec} &\multirow{3}{*}{ML-1M} & Oracle & 3.63 & 12.33 & 19.82 & 30.34 & \textbf{60.56} & 3.63 & 7.97 & 10.38 & 13.03 & 18.56 \\
  && DWRS-D & \textbf{3.70} & 12.36 & 19.82 & 30.34 & 60.21 & \textbf{3.70} & 8.01 & 10.42 & 13.06 & 18.53 \\
  && DWRS-U & 3.64 & \textbf{12.55} & \textbf{20.01} & \textbf{30.40} & 60.43 & 3.64 & \textbf{8.09} & \textbf{10.49} & \textbf{13.11} & \textbf{18.62} \\
 \cline{2-13}
 &\multirow{3}{*}{Beauty} & Oracle & \textbf{0.45} & \textbf{1.84} & \textbf{3.13} & 4.94 & 12.03 & \textbf{0.45} & \textbf{1.14} & \textbf{1.55} & \textbf{2.01} & \textbf{3.28} \\
  && DWRS-D & 0.44 & 1.80 & 3.04 & \textbf{4.96} & \textbf{12.15} & 0.44 & 1.11 & 1.51 & 1.99 & 3.23 \\
    && DWRS-U & 0.44 & 1.69 & 2.93 & 4.88 & 11.99 & 0.44 & 1.06 & 1.46 & 1.95 & 3.23 \\
 \cline{2-13}
 &\multirow{3}{*}{Steam} & Oracle & \textbf{12.08} & 16.23 & 19.61 & 24.45 & 42.93 & \textbf{12.08} & \textbf{14.17} & \textbf{15.26} & 16.47 & 19.79 \\
  && DWRS-D & 12.03 & 16.17 & 19.52 & 24.34 & 42.77 & 12.03 & 14.12 & 15.19 & 16.40 & 19.71 \\
  && DWRS-U & \textbf{12.08} & \textbf{16.24} & \textbf{19.63} & \textbf{24.50} & \textbf{43.01} & \textbf{12.08} & \textbf{14.17} & \textbf{15.26} & \textbf{16.48} & \textbf{19.81} \\
 \hline
 \multirow{9}{*}{NARM} &\multirow{3}{*}{ML-1M} & Oracle & \textbf{7.28} & 19.39 & 27.24 & \textbf{37.74} & \textbf{64.08} & \textbf{7.28} & \textbf{13.44} & \textbf{15.96} & \textbf{18.61} & \textbf{23.43} \\
  && DWRS-D & 7.04 & \textbf{19.49} & \textbf{27.36} & 37.38 & 63.78 & 7.04 & 13.41 & 15.94 & 18.46 & 23.32 \\
  && DWRS-U & 7.05 & 19.10 & 27.14 & 37.15 & 63.76 & 7.05 & 13.23 & 15.81 & 18.33 & 23.22 \\
 \cline{2-13}
 &\multirow{3}{*}{Beauty} & Oracle & 0.76 & 2.27 & 3.49 & 5.17 & 11.12 & 0.76 & 1.52 & 1.91 & 2.33 & 3.41 \\
  && DWRS-D & 0.76 & \textbf{2.37} & \textbf{3.64} & 5.46 & 11.98 & 0.76 & \textbf{1.57} & \textbf{1.98} & 2.43 & 3.61 \\
    && DWRS-U & \textbf{0.78} & 2.32 & 3.61 & \textbf{5.56} & \textbf{12.56} & \textbf{0.78} & 1.56 & 1.97 & \textbf{2.47} & \textbf{3.73} \\
 \cline{2-13}
 &\multirow{3}{*}{Steam} & Oracle & 12.10 & 16.71 & 20.30 & \textbf{25.29} & \textbf{44.01} & 12.10 & 14.44 & 15.60 & 16.85 & 20.21 \\
  && DWRS-D & \textbf{12.15} & \textbf{16.74} & \textbf{20.31} & 25.27 & 43.99 & \textbf{12.15} & \textbf{14.48} & \textbf{15.62} & \textbf{16.87} & \textbf{20.24} \\
  && DWRS-U & 12.11 & 16.69 & 20.25 & 25.20 & 43.84 & 12.11 & 14.43 & 15.58 & 16.82 & 20.17 \\
 \hline
   \multirow{9}{*}{CL4SRec} &\multirow{3}{*}{ML-1M} & Oracle & 4.71 & 16.67 & 25.27 & \textbf{36.42} & \textbf{63.65} & 4.71 & 10.78 & 13.55 & 16.36 & 21.35 \\
  & & DWRS-D & \textbf{4.94} & \textbf{16.73} & \textbf{25.38} & 36.25 & 63.42 & \textbf{4.94} & \textbf{10.91} & \textbf{13.69} & \textbf{16.43} & \textbf{21.41} \\
  & & DWRS-U & 4.79 & 16.46 & 25.03 & 35.97 & 63.26 & 4.79 & 10.72 & 13.47 & 16.23 & 21.25 \\
 \cline{2-13}
 & \multirow{3}{*}{Beauty} & Oracle & 0.50 & 2.63 & 4.25 & 6.25 & 12.79 & 0.50 & 1.58 & 2.10 & 2.61 & 3.79 \\
  & & DWRS-D & 0.53 & 2.71 & 4.25 & 6.25 & 12.74 & 0.53 & 1.64 & 2.14 & 2.64 & 3.81 \\
  & & DWRS-U & \textbf{0.57} & \textbf{2.86} & \textbf{4.46} & \textbf{6.31} & \textbf{12.90} & \textbf{0.57} & \textbf{1.72} & \textbf{2.24} & \textbf{2.70} & \textbf{3.89} \\
 \cline{2-13}
 & \multirow{3}{*}{Steam} & Oracle & 11.97 & 16.90 & 20.70 & \textbf{25.94} & 45.31 & 11.97 & 14.47 & 15.69 & 17.01 & 20.49 \\
  & & DWRS-D & 11.95 & 16.88 & 20.63 & 25.85 & 45.29 & 11.95 & 14.45 & 15.65 & 16.96 & 20.46 \\
  & & DWRS-U & \textbf{12.00} & \textbf{16.96} & \textbf{20.71} & \textbf{25.94} & \textbf{45.38} & \textbf{12.00} & \textbf{14.52} & \textbf{15.72} & \textbf{17.04} & \textbf{20.53} \\
 \hline
  \multirow{9}{*}{DuoRec} &\multirow{3}{*}{ML-1M} & Oracle & 5.66 & 17.92 & 25.97 & 36.54 & 63.11 & 5.66 & 11.89 & 14.48 & 17.14 & 22.01 \\
  && DWRS-D & \textbf{5.91} & 17.86 & \textbf{26.35} & 36.77 & \textbf{63.13} & \textbf{5.91} & \textbf{11.99} & \textbf{14.72} & \textbf{17.35} & \textbf{22.19} \\
  && DWRS-U & 5.56 & \textbf{17.95} & 25.95 & \textbf{36.80} & 63.09 & 5.56 & 11.86 & 14.43 & 17.18 & 22.00 \\
 \cline{2-13}
 &\multirow{3}{*}{Beauty} & Oracle & 0.88 & 2.86 & 4.21 & 5.87 & 11.41 & 0.88 & 1.90 & \textbf{2.33} & \textbf{2.75} & 3.75 \\
  && DWRS-D & \textbf{0.92} & 2.86 & 4.17 & 5.83 & 11.29 & \textbf{0.92} & \textbf{1.91} & \textbf{2.33} & 2.74 & 3.73 \\
  && DWRS-U & 0.82 & \textbf{2.90} & \textbf{4.29} & \textbf{5.94} & \textbf{11.67} & 0.82 & 1.89 & \textbf{2.33} & \textbf{2.75} & \textbf{3.78} \\
 \cline{2-13}
 &\multirow{3}{*}{Steam} & Oracle & 12.15 & 16.96 & 20.61 & 25.74 & 44.79 & 12.15 & 14.59 & 15.77 & 17.05 & 20.47 \\
  && DWRS-D & 12.13 & 17.03 & 20.72 & 25.88 & 44.93 & 12.13 & 14.62 & 15.81 & 17.11 & 20.53 \\
  && DWRS-U & \textbf{12.20} & \textbf{17.05} & \textbf{20.78} & \textbf{25.90} & \textbf{44.99} & \textbf{12.20} & \textbf{14.66} & \textbf{15.86} & \textbf{17.14} & \textbf{20.57} \\
 \hline
\end{tabular}
\label{table:DWRS-D all model utility}
\end{center}
\end{table*}

\end{document}